Modelling and experimental verification of photoelectrical response of NV diamond spin centres.

*Josef Soucek[a,b], Michael Petrov[a,c], Michal Gulka[c,d], Emilie Bourgeois[a,c], Milos Nesladek[a,b,c]*

Corresponding author: Josef Soucek (josef.soucek@uhasselt.be)

[a]Institute for Materials Research (IMO), Hasselt University, Wetenschapspark 1, B-3590 Diepenbeek, Belgium.
[b]Faculty of Biomedical Engineering, Czech Technical University in Prague, Sítna sq. 3105, 27201 Kladno, Czech Republic.
[c]IMOMEC division, IMEC, Wetenschapspark 1, B-3590 Diepenbeek, Belgium.
[d]Institute of Organic Chemistry and Biochemistry of the CAS, Flemingovo nam. 2, Prague, Czechia.

## 1. Abstract

We report on a mathematical model of the photoelectric response of NV colour centres in diamond, that can be employed for sensing and quantum science information applications. Although the model applies to NV centre in diamond, it can be applied with small modifications to other semiconducting solid state qubits. In our model, we include the drift and collection of charge carriers as well as the presence of other defects via generation and recombination dynamics. Though the photoluminescence readout and the associated dynamics of the NV defect has been extensively studied experimentally and theoretically, so far, there has been no precise model for photocurrent readout, including these effects. In our description, we use a multilevel-level system including $m_S$=0, $m_S$=±1 ground and excited states, singlet state and the $NV^0$ neutral state. Also, the presence of substitutional nitrogen ($N_S$), which for example determines the spin coherence via the paramagnetic spin bath, is discussed together with presence of acceptor defects. We model the time-dependent occupation of all electronic sublevels and also consider the electronic charge transport from the Boltzmann transport equation, leading to information about the charge state transitions and recombination dynamics. ODMR and PDMR response as well as their quantum efficiencies, are calculated. On this basis, we determine an optimal parameter space for qubit operations, including the highest spin contrast and especially relate those to $N_S$ presence. The model is confirmed experimentally and can become a useful tool for optimisation of the performance of NV qubit photoelectric readout.

## 2. Introduction

An important category of quantum devices are solid-state spin qubits,[1] and in particular, the nitrogen-vacancy (NV) centres in diamond[2–6]. The electron spin of the negatively charged NV centre can be initialised and read based on spin-conserving optical transitions. Understanding these transitions is essential to precisely drive and control NV spins, and reach a high spin contrast and high quantum gate fidelities. The model for $NV^-$ centre (i.e. negatively charged NV centre) optical transitions consists of ground and excited spin triplet levels and the spin singlet states and is very well developed[7–9,10], even though there are still some ongoing discussions about the metastable singlet state relaxation.[11–13] For example, Tetienne et.al. claim the probability of the relaxation from the metastable singlet state to the $m_s$ = 0 state of being 70-50%[12], while Wirtitsch et.al. give the value of 90%[13]. Some models that include NV centre ionisation transitions also exist, however for some transitions, the specific rates are still unknown[13,14]. NV centres ionisation is especially important in the photoelectric detection of magnetic resonance (PDMR), which we have proposed as an alternative method to the optical detection of magnetic resonances (ODMR)[15,16]. PDMR measurement requires collection of the charges generated by NV centres[17,18], which are transported through diamond under an external electric field and collected by the electrodes deposited on the diamond surface[18], as depicted in Figure 1. The PDMR method presents several advantages, such as easier device integration, higher detection rates that result in a high signal-to-noise ratio (SNR), and spatial resolution that is independent of the confocal diffraction limit and can be miniaturised, since it depends mainly on the device size[15,17,19–23]. Spatial resolution plays a significant role in the fabrication of sensing devices and devices for quantum information science. The modelling of electron spin transitions can thus be used to design optimal devices and protocols and for the precise control of the NV qubit states, leading to an improved performance. It will also allow the design of optimised diamond materials and devise the influence of impurities on the photoelectric signal.

Unlike ODMR, when describing PDMR, one has to consider the electronic transport and charge recombination dynamics. Here and further we call recombination any process that results in a loss of at least one free charge carrier from either band. Understanding of these effects is necessary to properly interpret the NV photocurrent response. The additional challenge to consider is photoionisation of the other defects. Certain defects can act as photoionisation recombination centres for charge carriers, and they can significantly influence the PDMR spin contrast and the detection rate[24–27]. These effects have not been included in modelling so far and represent a rather complex mathematical description.

Here, we develop a mathematical model of the ODMR/PDMR readout, including the transport and charge carrier recombination dynamics, thus enriching the currently existing models by these effects. That allows us to calculate the time-dependent occupation of all electronic sublevels. Another important issue is the presence of defects in the diamond lattice, in particular the substitutional nitrogen, i.e. P1-EPR centre (denoted further as $N_S$). $N_s$ is known to cause a paramagnetic spin bath, that is responsible for the NV spin decoherence. Therefore, methodologies for easy evaluation of the $N_s$ impact are of a high interest. Further on, the $N_s$ also influences the PDMR spin contrast via parasitic photocurrent generation[15,18] and the exact mechanism responsible for it is still unknown. Also, the presence of other defects in diamond, such as an acceptor defect level, which we introduced based on our recent experimental observations[28], influences the PDMR readout and is studied here.

Our model is based on six energy sublevels of $NV^-$ ($m_S$ = 0 and $m_S$ = ± 1 ground and excited state triplet levels, $NV^-$ singlet state and $NV^0$ charge state level), as well as the energy levels associated to $N_S$ and to the acceptor defect mentioned above. The system is solved numerically, as detailed in Supplementary Information. The model allows us to predict the optical or photoelectrical spin contrast as a function of the laser power. It also enables the evaluation of the NV centre quantum efficiency for photoluminescence (PL) photons and for electron-hole pair generation (photocurrent) and their dependence on the $N_s$ and the acceptor defect. Also, our results show that for higher incoming laser power, the quantum efficiency for electrical readout exceeds the quantum efficiency for optical detection. Finally, we discuss the differences in the photoluminescence and photocurrent response as a function of the laser power, from which we estimate the concentration of $N_S$ defects that interact with the electronic charge during the transport. Experimental verification shows an excellent agreement with the theory.

# 3. Theoretical model

## 3.1. NV electronic state transitions

First, we concentrate on the solution of the $NV^0$ / $NV^-$ rate system in the presence of charge carrier drift. In particular, the nitrogen-vacancy defect has three charge states. The neutral ($NV^0$) and the negative ($NV^-$) charge states are bright in photoluminescence, whereas the positive ($NV^+$) does not lead to any luminescent transitions[29]. Here, we consider only the $NV^0$–$NV^-$ conversion used in the spin electrical readout[4], since the $NV^+$ is not directly involved in these transitions in first approximation[15]. The model is depicted in Figure 1A.

The $NV^-$ energy level structure is represented here by five energy levels. The $m_s = 0$ and unsplit $m_s = \pm 1$ are ground (GS) and excited (ES) triplet states (respectively denoted as $^3A_2$ and $^3E$ ), while the metastable spin singlet state is denoted as $^1E$ (here for simplicity the singlet states are depicted as one level, justified by the $^1A_1$ singlet excited state short relaxation time). The ground and excited states of $NV^0$ (respectively noted $^2E$ and $^2A_1$) are doublet spin ½ systems. An electron can be photoionised from the valence band (VB) to the $NV^0$ state via a two-photon mechanism[14] or directly by a one-photon process, as proposed and calculated[30,31]. Following this back-conversion process, the $NV^0$ gains an extra electron from the valence band, and its charge state returns to $NV^-$. During this photoionisation, a hole charge carrier is produced in the valence band. The second part of the loop consists of the two-photon photoionisation of $NV^-$ from its ground state[15], during which a free electron is generated in the conduction band. Both the free holes and free electrons are transported under an applied electric field. The charge carrier drift and recombination are essential parameters in the whole process[14]. The charge carriers are then collected by the electrodes, and this process is completed with NV ending again in the $NV^0$ charge state.

Consequently, after photoionisation, the sum of the electron and hole currents is measurable by an external circuit. This current depends on the electric field, number of incoming photons, charge mobility and the recombination lifetime. The recombination lifetime is defined as the reciprocal value of the electron and hole recombination rates and is determined by the slowest of these two. In addition, the presence of various defects in diamond and their occupation can influence the value of the recombination lifetime significantly. This will affect the NV electron spin contrast, since it depends on the charge transition rates, from $NV^0$ to $NV^-$ and back.

The $NV^-$ ground state zero-field splitting (ZFS) is 2.87 GHz[32]. In the absence of Zeeman splitting, the application of microwaves at resonance results in a drop in PL as well as a drop in the photocurrent from the NV. The microwave resonance contrast is defined as the amplitude of the drop with respect to the total signal. The reason for the drop is the higher probability of transition from the excited triplet state of NV- to its metastable dark singlet state. At low optical excitation power, the ODMR optical contrast is provided by the ratio between the excited-to-ground luminescent triplet state transitions and the non-radiative shelving transitions via the metastable state singlet. However, when the laser excitation power is increased, $NV^-$ depopulation occurs by two-photon ionisation (i.e. $NV^0$ generation), which is then followed by a back conversion to $NV^-$, involving electron from the valence band. This process plays an important role in determining the spin contrast[13]. As PDMR operates in such a regime, the back conversion is crucial because, during this transition, the $^1E$ metastable singlet state is directly populated from the $NV^0$ excited state[13]. This process can be significantly enhanced by very high incoming photon rates[13], which were used recently for designing tailored 2-pulse ODMR protocols leading to detected optical spin contrast of above 46%[13]. These transitions are depicted in Fig. 1A, considering the excitation rate $k_1$ and the ionisation rate $ion_1$, representing the two-photon ionisation process of the $NV^-$ charge state. Rate $ion_2$ describes the photoionisation of $NV^0$ from the valence band. The recombination rates can be either radiative ($k_2$ - here, for simplicity, we consider that, due their short lifetimes, the $m_s = 0$ and $m_s = \pm 1$ spin sublevels have identical radiative recombination rates) or non-radiative ($k_{3\text{-}6}$). The last group of recombination paths relates to transitions involving the conduction or the valence bands, in particular, the recombination rate of electrons from $NV^-$ ground state to the valence band ($rec_1$) or the recombination of the electron from the conduction band to $NV^0$ ($rec_2$)[25], or recombination via defect levels, are discussed later. Transition between the sublevels of $NV^-$ ground state ($m_S= 0$ and $m_S= \pm1$) is represented by the transition with rate $k_{MW}$ (Supplementary info - Eqn. S1). Details about the rates and their values can be found in Supplementary info, Table S1.

The transition rates between the GS, ES, and metastable state are well-known at this point. However, as mentioned above, there is still an ongoing discussion concerning the relaxation rates from the metastable state. Robledo et.al.[11] and Tetienne et al.[12] claim that the probability of the relaxation to the $m_s$ = 0 state is around 50%-70%, while Wirtitsch et al.[13] give the value of 90%, which is in agreement with ab initio calculations by Thiering et al.[33] This is a crucial point, since this greatly affects the initialisation of the electron spin, which in turn affects the ODMR and PDMR spin contrasts. The key difference between the two methods used in these studies was that the former results are derived from the modelling of the time trace of the fluorescence response after a ~ 50 ps excitation pulse (Tetienne[12]), and the latter study is based on the response to a longer excitation pulse (~1 μs, Wirtisch[13]). This creates an open question if the difference in measuring methodologies might be playing a role in this apparent discrepancy. However, a recent study by Ernst et al.[34] investigates the fluorescence response after long excitation pulses similar to the study by Wirtitsch et al.[13], and yet the results are in agreement with those by Tetienne et al.[12] Our model can therefore shine light on this question.

The ionisation cross-section for the NV charge transitions is another parameter that is under debate. Some of the ionisation rates were estimated by modelling and fitting to the experimentally determined photoluminescence rate and photocurrent dependence on the laser power[35]. For example, $ion_1$ rate is not exactly known. It has been suggested by Wirtisch[13] that this rate was about 23% of $k_1$. In our case we started with this value and by modelling experimental ODMR traces for higher power we found the best agreement for $ion_1$ equal to 25% of $k_1$. More detail on this procedure and the exact value of the modelled rates can be found in Supplementary Information (SI).

From our theoretical model, we are, in addition, able to calculate the quantum efficiency (QE) as the sum of all the detectable photon or electron-hole pairs, involved in the transitions, weighted by the number of incoming photons. Therefore, the optically and electrically detected signals are separable and can be traced individually. Both depend differently on the incoming photon flux via different rate coefficients. In addition, for PDMR readout, the QE is influenced by charge carrier recombination rates as it relates to the number of electron-hole pairs that reach the electrodes and are collected at that point. For this reason, the QE is also influenced by the presence of various defect levels in the bandgap of diamond, in particular $N_S$, which serve as recombination centres and traps. Those can significantly modify the recombination dynamics and are also reflected in the QE power dependence, as studied below . It is also, in principle, possible to have unipolar current through charge injection from an electrode[24], however the current we observe without laser illumination is negligible compared to photocurrent, therefore we do not consider it.

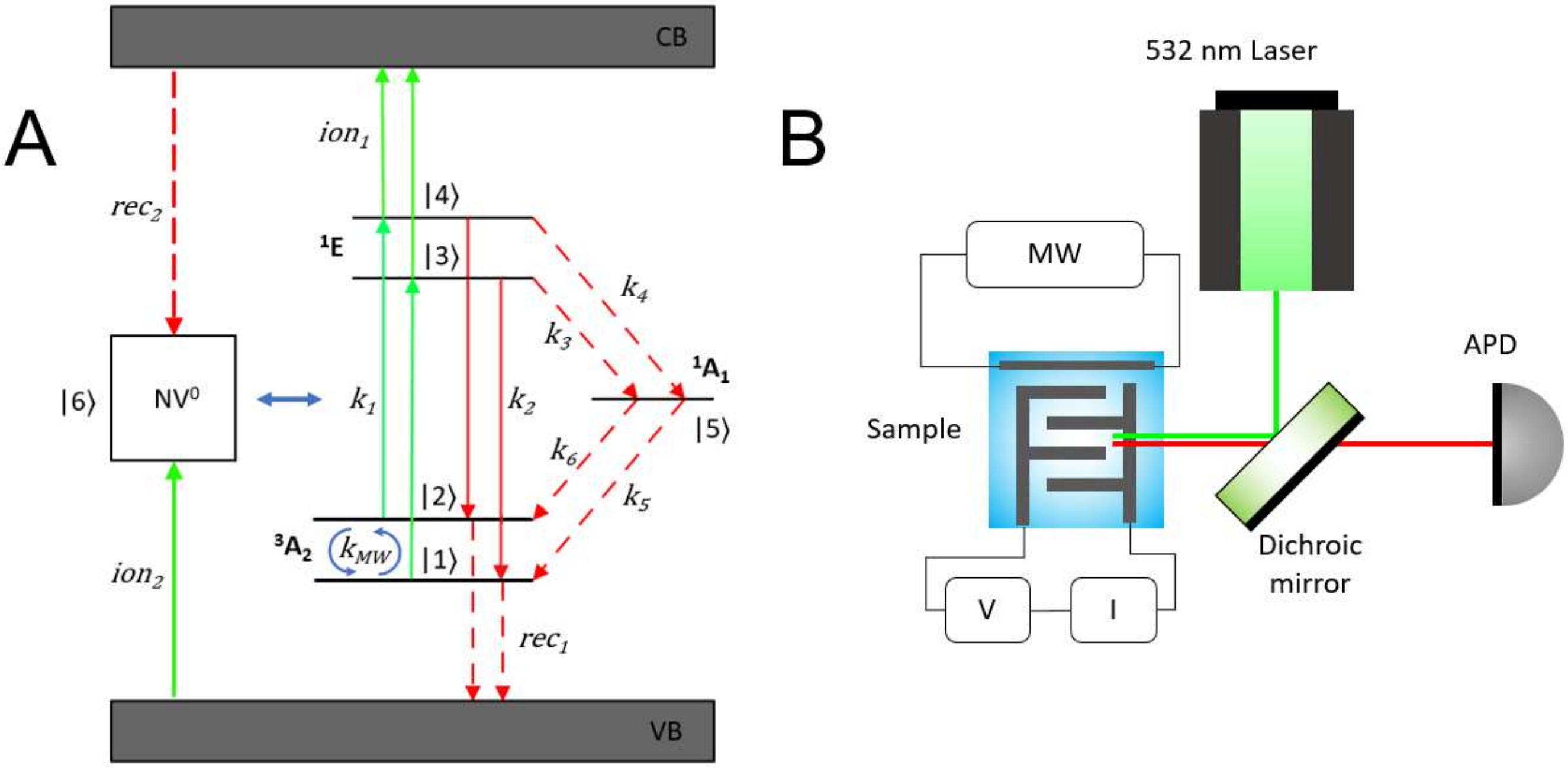

**Figure 1A-B**: Figure 1A shows the energy level system of NV defect. Levels 1-2 represent the GS of $NV^-$, 3-4 the ES of $NV^-$, 5 the metastable state (MS), 6 denotes the $NV^0$ charge state, VB denotes the valence band and CB the conduction band. The colour of the arrows defines excitation (green), relaxation (red), change in the charge state or ground state sublevel (blue). The solid lines relate to ionisation or radiative relaxation, and the dashed lines relate to non-radiative relaxation. $k_i$ denotes the following transitions between energy levels of $NV^-$ charge state: $k_1$ (GS to ES), $k_2$ (ES to GS), $k_3$ (ES $m_S$= 0 to MS), $k_4$ (ES $m_S$=±1 to MS), $k_5$ (MS to GS $m_S$=0), $k_6$ (MS to GS $m_S$= ±1), $k_{MW}$ (GS $m_S$= 0 to GS $m_S$= ±1 and back). $ion_1$ denotes the ionisation transition from ES of $NV^-$ to CB, $ion_2$ the ionisation transition from VB to $NV^0$ charge state, $rec_1$ stands for the recombination between GS of $NV^-$ and VB, $rec_2$ for the recombination between the CB and $NV^0$. Figure 1B shows a simplified scheme of the confocal setup used for experimental confirmation of the model. A microwave field is applied to a metal strip fabricated on the diamond surface; a constant voltage is applied between the electrodes. Laser light is focused on an NV centre between the electrodes, NV photoluminescence is collected by an avalanche photodiode (APD), and the current between electrodes is measured with an amperemeter. To filter out some of the noise and to improve the signal-to-noise ratio, we use a lock-in amplifier for photocurrent detection.

As an initial condition, we arbitrarily set the NV centre into the neutral $NV^0$ state, and we mark the occupation probability of level 6 as equal to 1 (i.e., the NV centre being in the neutral state is denoted as p =1 at the time t = 0, i.e., $p_{|6\rangle,t(0)}=1$). After the incoming photon impinges on the NV centre, the electron is excited from $E_v$ via a two-stage process[14] or a single-stage process[32], forming the $NV^-$ charge state. In this model, for simplicity, only one-photon (effective) back-conversion processes are considered. The excitation and ionisation rates ($k_1$,$ion_i$) are laser dependent, and they can be described, for example of rate $k_1$, as $k_1 = W_{k1} \cdot P_{laser}$ where $W_{k1}$ is the optical pumping parameter and $P_{laser}$ is the laser power in mW. More information about the calculation of each rate can be found in SI. The dynamics of the electron transitions can be described mathematically by the following set of differential equations.

$$\frac{p_{|1\rangle}}{dt} = -k_1 \cdot p_{|1\rangle} + k_2 \cdot p_{|3\rangle} + k_5 \cdot p_{|5\rangle} + A_k \cdot ion_2 \cdot p_{|6\rangle} \cdot p_{|VB\rangle} - k_{MW} \cdot p_{|1\rangle} + k_{MW} \cdot p_{|2\rangle} + A_k \cdot rec_2 \cdot p_{|6\rangle} \cdot p_{|CB\rangle} - D_k \cdot rec_1 \cdot p_{|1\rangle} \cdot (1 - p_{|VB\rangle}) \quad (1)$$

$$\frac{p_{|2\rangle}}{dt} = -k_1 \cdot p_{|2\rangle} + k_2 \cdot p_{|4\rangle} + k_6 \cdot p_{|5\rangle} + k_{MW} \cdot p_{|1\rangle} - k_{MW} \cdot p_{|2\rangle} + B_k \cdot rec_2 \cdot p_{|6\rangle} \cdot p_{|CB\rangle}$$

$$-E_k \cdot rec_1 \cdot p_{|2\rangle} \cdot (1 - p_{|VB\rangle}) + B_k \cdot ion_2 \cdot p_{|6\rangle} \cdot p_{|VB\rangle} \quad (2)$$

$$\frac{p_{|3\rangle}}{dt} = k_1 \cdot p_{|1\rangle} - (k_2 + ion_1 + k_3) \cdot p_{|3\rangle} \quad (3)$$

$$\frac{p_{|4\rangle}}{dt} = k_1 \cdot p_{|2\rangle} - (k_2 + ion_1 + k_4) \cdot p_{|4\rangle} \quad (4)$$

$$\frac{p_{|5\rangle}}{dt} = k_3 \cdot p_{|3\rangle} + k_4 \cdot p_{|4\rangle} - (k_5 + k_6) \cdot p_{|5\rangle} + C_k \cdot ion_2 \cdot p_{|6\rangle} \cdot p_{|VB\rangle} + C_k \cdot rec_2 \cdot p_{|6\rangle} \cdot p_{|CB\rangle} \quad (5)$$

$$\frac{p_{|6\rangle}}{dt} = -ion_2 \cdot p_{|6\rangle} \cdot p_{|VB\rangle} + ion_1 \cdot (p_{|4\rangle} + p_{|3\rangle}) - rec_2 \cdot p_{|6\rangle} \cdot p_{|CB\rangle}$$

$$+rec_1 \cdot p_{|1\rangle} \cdot (1 - p_{|VB\rangle}) \quad (6)$$

**Equations 1-6:** The set of equations for the electronic transitions on the NV defect. Eqn. 1 represents the rate equations for the $m_S$= 0 spin sublevel of $NV^-$ GS, Eqn. 2 for the $m_S$= ±1 spin sublevel of $NV^-$ GS, Eqn. 3 for the $m_S$= 0 spin sublevel of $NV^-$ES, Eqn. 4 for the $m_S$= ±1 spin sublevel of $NV^0$ ES, Eqn. 5 for the metastable singlet state of $NV^-$, Eqn. 6 for $NV^0$ charge state. Here p stands for the population of each specific energy sublevel (see Fig. 1), *CB* and *VB* stand for Conduction and Valence bands, *rec* stands for recombination rates, *ion* for ionisation rates and $A_k$-$E_k$ stand for the recombination coefficients[13] see Supplementary Information, SI, Table S1).

$$\frac{p_{|CB\rangle}}{dt} = -rec_2 \cdot p_{|6\rangle} \cdot p_{|CB\rangle} + ion_1 \cdot (p_{|4\rangle} + p_{|3\rangle}) + (1/e) \cdot \nabla j_{CB} \quad (7)$$

$$\frac{p_{|VB\rangle}}{dt} = -ion_2 \cdot p_{|6\rangle} \cdot p_{|VB\rangle} + rec_1 \cdot p_{|1\rangle} \cdot (1 - p_{|VB\rangle}) + (1/e) \cdot \nabla j_{VB} \quad (8)$$

**Equations 7-8**: Transport and transition rate equations involving the valence and conduction band electron and hole population; $p_{CB}$ stands for the population of the conduction band, and $p_{VB}$ stands for the population of the valence band, *e* stands for the electric charge of the electron. $\nabla j_{CB}$ and $\nabla j_{VB}$ are the divergence of the drift currents.

Rate equations 1-6 use the rate constants listed in Supplementary information based on available data[12]. The recombination coefficients ($A_k$-$E_k$) are used to represent the percentage distribution of the known recombination fluxes to different energy levels according to the Equations 1-6.[13] An example is the ionisation flux from $NV^0$ to $NV^-$ , which uses the recombination coefficients Ak ($NV^0$ to $NV^-$GS $m_S$= 0), Bk ($NV^0$ to $NV^-$GS $m_S$= ±1), and Ck ($NV^0$ to $NV^-$GS metastable singlet), where the sum of these recombination coefficients is equal to us 1 (total flux from $NV^0$ to $NV^-$). The equation 8 contains the $p_{VB}$ element.. $P_{VB}$ represents the electron available in the valence band for ionisation ($p_{VB}$=1 electron available for ionisation, $p_{VB}$=0 hole available for recombination). Equations 7-8 also contain the divergence of the drift currents $\nabla j_{CB}$ and $\nabla j_{VB}$. These elements are discussed in section 3.4.

The photoluminescence signal $I_f$ of $NV^-$ can be evaluated from the radiative relaxation fluxes from the ES to the GS of $NV^-$ [12]. Here $p_3$ (Eqn. 3) and $p_4$ (Eqn. 4) are the populations of the ES of $NV^-$ and $k_2$ is the rate of radiative transitions from ES to GS of $NV^-$

$$I_f = k_2 \cdot (p_{|3\rangle} + p_{|4\rangle}) \quad (9)$$

The steady-state photocurrent $I_p$ is described as the sum of electron $I_e$ and hole $I_h$ currents, leading to a total photocurrent $I_p = I_e + I_h$. Each of the currents is defined as [36]

$$I_e = e \cdot G_e \cdot \tau_e \cdot \mu_e \cdot E \quad (10)$$

$$I_h = e \cdot G_h \cdot \tau_h \cdot \mu_h \cdot E \quad (11)$$

Where $e$ is the electric charge of the electron, $E$ is the applied external electrical field, $\mu_e$ and $\mu_h$ are the mobility of the electron and the hole charge carriers, $\tau_e$ and $\tau_h$ are their recombination lifetime, and $G_e$ and $G_h$ are the photogeneration rates for electrons and holes. The details of the generation rate and carrier lifetime calculations can be found in Supplementary Information (SI). Results and discussion of the quantum response of NV can be seen in section 4.1.

By calculating the total photocurrent as the sum of the electron and hole currents $I = I_e + I_h$ in the case of MW excitation off and on, we can calculate the total theoretical spin contrast of the system by the formula:

$$C = \frac{I_{off} - I_{on}}{I_{off}} \quad (12)$$

The quantum efficiency (QE) is then calculated as the ratio between photoluminescence rate and incoming photon rate ($QE_f$) or between generated charge carrier collection rate and incoming photon rates in the case of photocurrent $QE_p$ by the formula:

$$QE_f = \frac{I_f}{k_1(p_{|1\rangle} + p_{|2\rangle}) + G_e + G_h} \quad (13)$$

$$QE_p = \frac{G_e + G_h}{k_1(p_{|1\rangle} + p_{|2\rangle}) + G_e + G_h} \quad (14)$$

**Equations 13-14**: Quantum efficiency equations for the photocurrent and photoluminescence.

Results and discussion of the quantum response of NV can be seen in section 4.1.

### 3.2. Additional defects - $N_s$ defect

Diamond crystals contain several types of point defects. The substitutional nitrogen $N_S$ is one of the most common defects in diamonds[37–39]. During diamond fabrication by chemical vapor deposition (CVD), only typically of about 1% of $N_s$ is converted to NV, inevitably causing $N_S$ concentration to exceed that of NV by a factor of 100 or more[40,41]. A similar situation occurs during nitrogen implantation, with a formation yield typically below 10% for shallow NV centres, though recently, the implantation efficiency has been increased by co-doping, leading to NV formation yield as high as 75%[42]. Thus, when considering the photoionisation and recombination processes contributing to the photocurrent, one needs to count with the photo ionisation of the $N_S$ defect and the recombination of free carriers on this defect. In PDMR, $N_S$ photoionisation can cause a decrease in the total photocurrent by reducing the recombination lifetime of charge carriers and a decrease in the photoelectrically detected spin contrast.[18,22,30] The related electron transitions for neutral $N_s^0$ are simplified below (Fig. 2B). In reality, a relaxation of the defect energy level[43] occurs after the photoionisation of the neutral $N_S^0$, leading to $N_S^+$ making the $N_S$ defect level relax in energy to about of 4.6 eV[44] above the valence band maximum. Consequently, the photo-induced back-conversion from $Ns^+$ to $N_S^0$ is impossible under green 532nm illumination. The $N_S^+$ can be back-converted to $N_s^0$ by recombination of the free electrons from the conduction band. When $N_s$ defect is included, it is necessary to implement an extra equation into the system.

$$\frac{p_{|8\rangle}}{dt} = -ion_3 \cdot p_{|8\rangle} + rec_3 \cdot \left(1 - p_{|8\rangle}\right) \cdot p_{|CB\rangle} - rec_4 \cdot p_{|8\rangle} \cdot (1 - p_{|VB\rangle}) \quad (15)$$

**Equation 15:** Rate equation for the population of the $N_S$ defect, where $p$ stands for the population of energy levels and indexes of $p$ stand for: 8 ($N_s^0$), CB (conduction band), VB (valence band), rec stands for recombination rates: $rec_3$ (CB to $N_s^0$) and $rec_4$ ($N_s^0$ to VB). Ionisation rate $ion_3$ stands for the ionisation rate from $N_s^0$ to CB. Graphical representation can be seen in Fig. 2B.

Results and discussion on the influence of Ns defects on the PDMR contrast can be found in section 4.2.

## 3.3. Acceptor defects

Even in the highest-purity CVD-diamond samples, in addition to $N_S$, several other types of defects can be present, for example, neutral and charged vacancies, di-vacancies, interstitials, and their complexes with N or hydrogen[6,45]. The influence of these defects on the optical and photoelectric spin contrast has not been studied in detail so far. Recently, our experimental results suggested the existence of such an acceptor level[28]. This defect level (noted level X) has the ability to trap an electron or to serve as a recombination centre (as depicted in Fig. 2A)[18,28]. We found earlier that the presence of acceptor defects could lead to intriguing behaviour, such as the inversion in the sign of the PDMR contrast[28]. Results and discussion on the influence of the acceptor defect level X on the PDMR contrast can be found in section 5.2. We have tentatively attributed this level to the single vacancy V[28].

In addition to $N_S$, we describe the transitions on the defect level X as an extra equation to the set (1-8):

$$\frac{p_{|7\rangle}}{dt} = \left(1 - p_{|7\rangle}\right) \cdot ion_4 \cdot p_{|VB\rangle} + rec_5 \cdot \left(1 - p_{|7\rangle}\right) \cdot p_{|CB\rangle} - rec_6 \cdot \left(1 - p_{|VB\rangle}\right) \cdot p_{|7\rangle} \quad (16)$$

**Equation 16**: Rate equation for X defect where p stands for the following levels' population: 7 ($X^0$), CB (conduction band), VB (valence band), *rec* stands for recombination rates: $rec_5$ (CB to $X^0$) and $rec_6$ ($X^0$ to VB). Ionisation rate $ion_4$ stands for the ionisation rate from VB to $X^0$. Graphical representation can be seen in Fig. 2A.

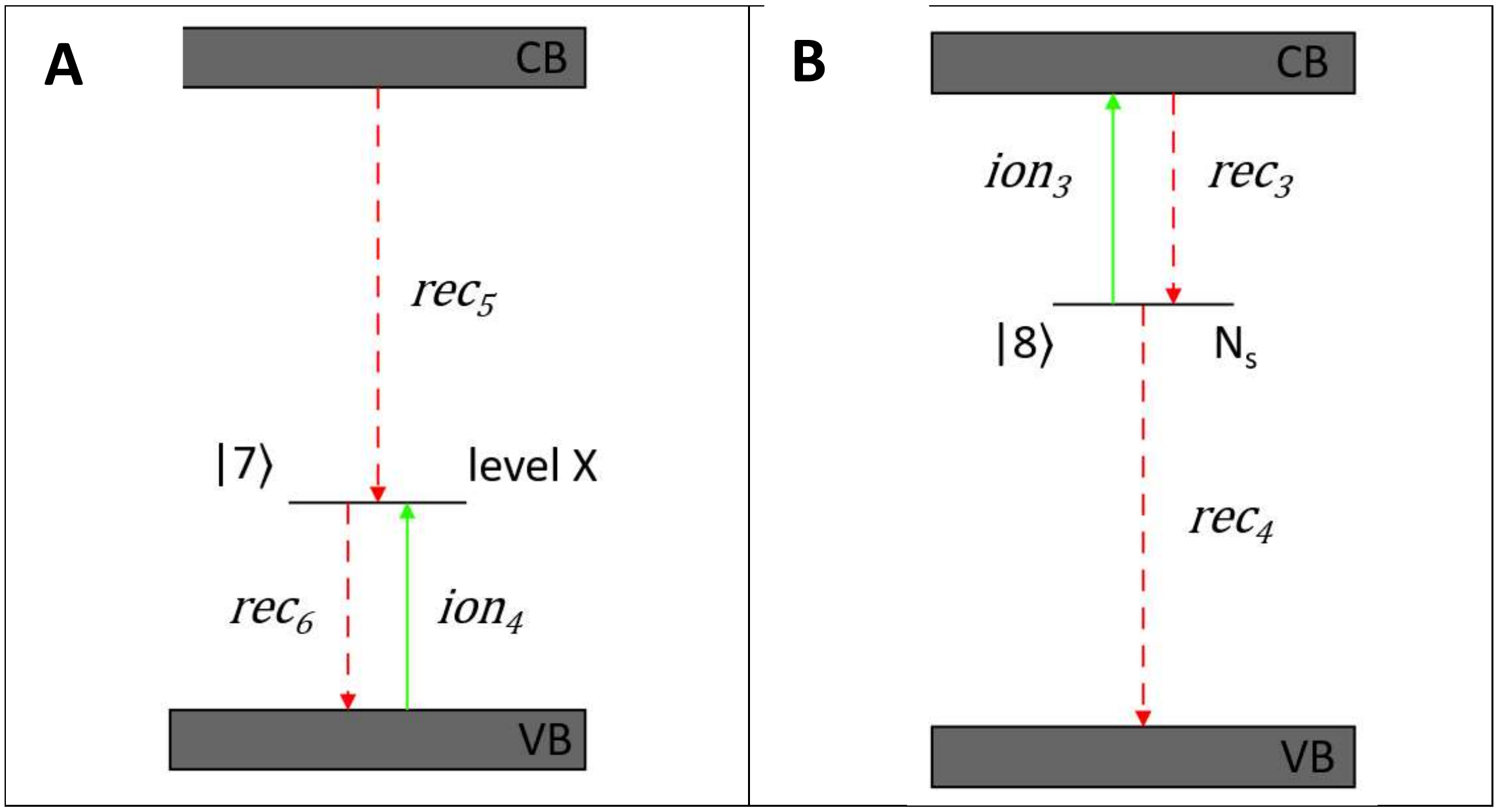

**Figure 2A-B:** Fig. 2A shows the proposed structure of the acceptor level X. $Rec_5$ is the recombination rate from the conduction band (CB) to X, $rec_6$ is the rate of recombination of the electron from X to the valence band (VB), rate $ion_4$ is the ionisation rate from VB to X. This ionisation rate produces a hole in the VB and the level is designed as a strong hole producer. Fig. 2B shows a model of the $N_s$ defect. The system is considered here as a one energy level system with recombination rates $rec_3$ and $rec_4$ and ionisation rate $ion_3$.

### 3.4. Charge carrier drift

The charge carrier drift is a crucial component of the PDMR scheme and also plays an important role in the recombination processes as well as charge carrier collection by electrodes[24]. The drift leads to redistribution of charge onto defects outside the illumination spot. By including it in the model, we account for recombination via the defects present on the way to the collecting electrode, which, as we show, have an effect on the photocurrent spin contrast. Distance of our electrodes is 5 µm and in this range we consider isotropic movement of the charge carriers. Our model is valid for small electrode distances, in the case of larger electrode gaps (more than 10-15µm), preferable charge carrier drift paths need to be considered[24]. In order to model the charge carrier drift, we represent the gap between the electrodes as a one-dimensional space, where the spatial coordinate is the current position of the transported electron with respect to the electrodes. This one-dimensional space is discretised into bins that allow numerical calculation over the spatial coordinate. Additional information about this process can be found below and in SI.

# 4. Methods

## 4.1 Experimental arrangement

To verify the model experimentally, we used an HPHT type-IIa diamond sample (New Diamond Technology) with $N_s$ concentration of 20 ppb and containing individual NV centres. The sample is equipped with Ti/Al electrodes separated by 5 µm and a microwave strip created by photolithography on the diamond surface. A single NV centre used for measurements is located at a depth of 0.1 µm below the diamond surface, 35 µm away from the microwave strip, 3.5 µm from one of the electrodes, and 1.5 µm from the other electrode. The sample is mounted on a custom-built confocal setup equipped with a microscope objective with a numerical aperture of 0.9 (Fig.1B). Laser light with 532 nm wavelength (Laser Quantum) is used for NV centre excitation and NV photoluminescence is collected, filtered out from the pump beam, and sent to an avalanche photodetector. The microwave strip is connected to an arbitrary waveform generator (Keysight AWG), which plays the role of a microwave source. Electrodes are connected to a voltage source and a current amplifier. The external current is detected with a lock-in-amplifier, functioning as an amperemeter (See Fig. 1B). To measure the magnetic resonance contrast, the microwave frequency is swept in a 100 MHz range around the resonance line of 2.87 GHz (Fig.4).

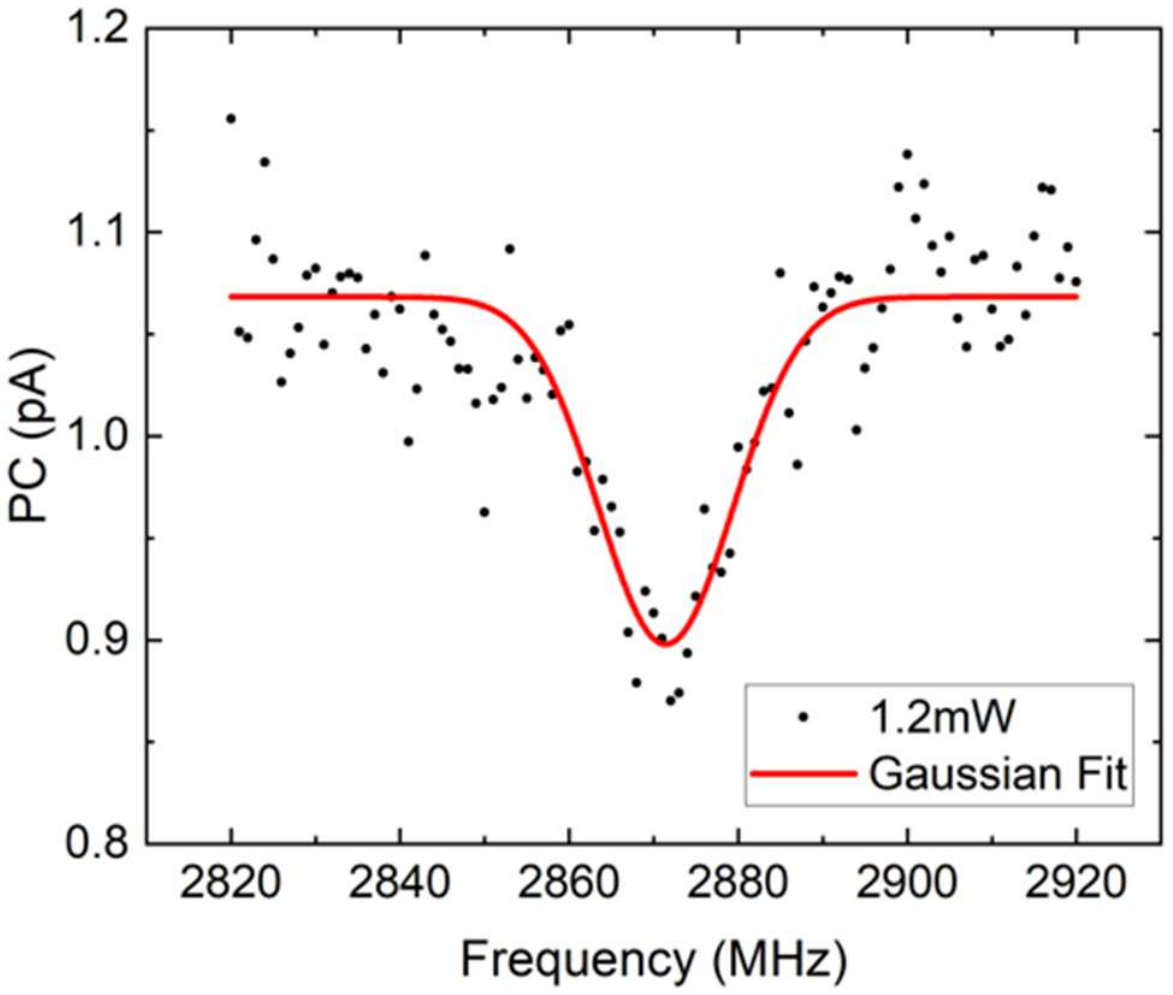

**Figure 4:** Example of a PDMR spectrum of a single NV defect of the used HPHT sample, for a laser power of 1.2 mW. The experimental trace is fitted by a Gaussian distribution.

### 4.2 Rate modelling

Modelling is performed using MatLab environment and objected-oriented programming (OOP). We developed a one-dimensional model in which we discretise the inter-electrode gap into small cells (bins). We position a single NV defect within the central bin of this mesh structure. We assume that the MW-induced magnetic field is perpendicular to the NV quantisation axis and the NV defect is in the $NV^0$ charge state at the time t=0.

The solution of the equation set is obtained through numerical integration of the rate equations (1-8) and (15-16). For each of the mesh cell element $N_i$ we solve the set of the rate equations. $N$ is set to 11 (which gives us a maximum total number of 110 equations), which was chosen as an optimum, taking into account the space resolution and the calculation time (for more information, see SI Appendix, Fig. S4A). We define a time 1D array with discretised time intervals. The equation set is then solved for each of the time bin, with the end-time vector $t_{end}$ equalling the time at which the system reaches the steady state. The outcome of the numerical integration process gives us a 3D matrix containing the population of each energy level for all time intervals. These results are used to calculate the total photocurrent and allow us to calculate the ODMR and PDMR contrast. More details about the model construction, initial condition and numerical integration can be found in Supplementary Information (SI).

## 5. Results

### 5.1 Charge dynamics of a single NV defect

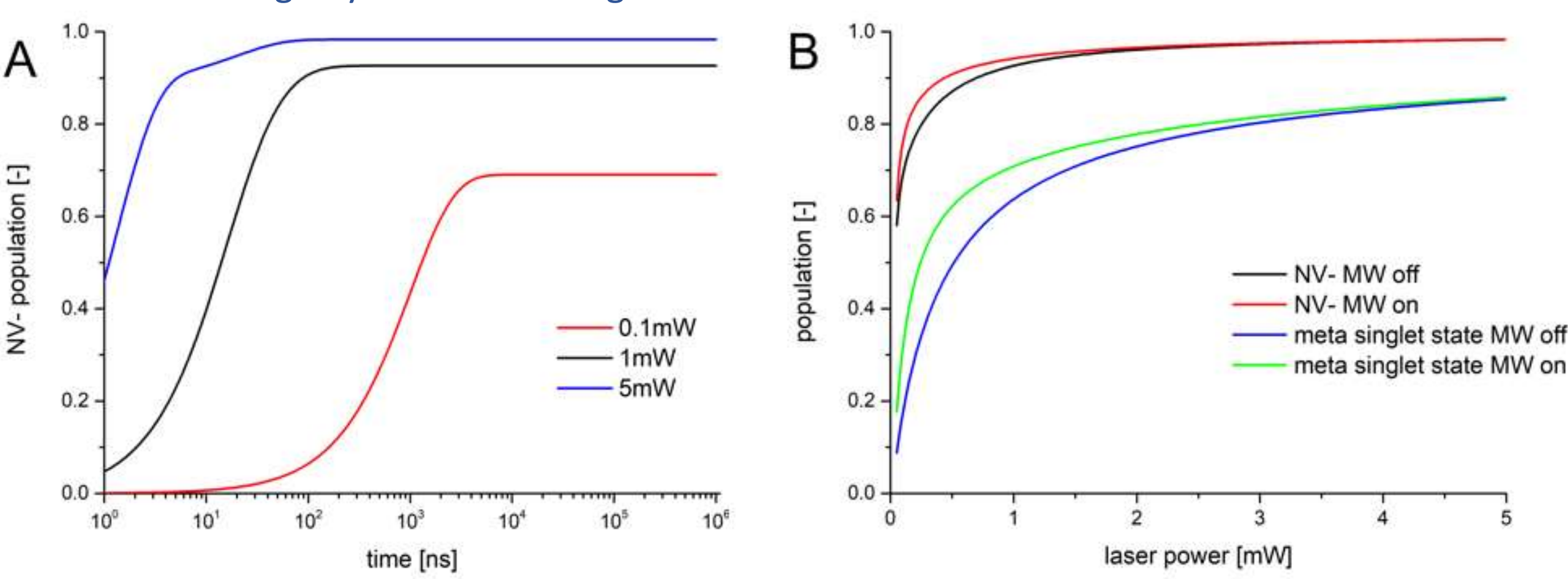

**Figures 5A-B:** Fig. 5A shows the time-dependent population of the $NV^-$ defect charge state without the presence of any other defects in the diamond crystal and without the resonant microwave field application. The system of partial differential equations is solved for three different laser powers (red – 0.1 mW, black – 1 mW, blue – 5 mW). Fig. 5B shows the laser power dependence of the population of the NV-charge state and the metastable state singlet at the steady state. The population of the $NV^-$ charge state is calculated with and without the resonant microwave field in black (MW off) and red colour (MW on), while the metastable singlet state population is represented in green (MW on) and blue colour (MW off).

Fig. 5A shows the calculated time-dependent photodynamic changes of NV- charge state occupation for 0.1 mW, 0.5 mW, and 5 mW laser power excitations. At the time 0, as defined previously, NV centre is in the $NV^0$ charge state. The first part of the trace shows an increasing population of the $NV^-$ charge state with elapsed time. A steady state is reached faster for the higher laser powers, and the population of the $NV^-$ charge state at equilibrium is increasing with the laser power as expected. Steady-state is reached at 5.6 µs for the laser power of 0.1 mW, 190 ns for the laser power of 1 mW, and 78 ns for the laser power of 5 mW with a charge state distribution ($NV^0/NV^-$) of ~31/69 % for 0.1mW, ~8/92 % for 1 mW and ~2/98 % for the 5 mW laser power.

When the laser power excitation is increased to 5 mW (Fig. 5A blue trace), a inflection point in $NV^-$ population can be seen at t = 6 ns. This is due to the saturation of the excited state of $NV^-$ whose population is rising faster as the population of the metastable singlet state. In this model we consider the effective rate for the ionisation of $NV^0$ to $NV^-$ ground state as rate $ion_2$ (defined at eqn. S5), corresponding of 80% charge transition from the $NV^0$ to $NV^-$ ground state and about 20% directly to the metastable state singlet[13]. After the metastable singlet state occupation reaches its maximum, which is given by its lifetime, the second saturation occurs. This is also in accordance with ref[13], where it is shown that at high laser powers, the charge is stored at the metastable singlet state. This is a consequence of the long-lived metastable singlet ground state with a lifetime of ~ 200 ns at room temperature[11]. The visibility of the second inflection point increases with the laser power, also in accordance with[15].Based on the knowledge of the population of each of the individual sublevels, we can calculate the spin contrast using equation eqn 12.

Fig. 5B shows the charge state populations of the $NV^-$ defect and of the metastable singlet state at steady state as a function of the laser power, with a resonant microwave field on or off. These traces are used for calculation of the spin-contrast in Fig 6. The population of the $NV^-$ charge state starts to saturate at 1mW and is followed by saturation of the metastable singlet state at 1.5 mW. The population of $NV^-$ reaches a value of approximately 0.92. Results of the model show that saturation of the population of the $NV^-$ charge state is linked with the saturation of the metastable singlet state, where most of the charge is stored.

### 5.2. ODMR and PDMR spin contrast in the presence of other defects

It was reported[15,18] that the $N_S$ centre plays a significant role in photocurrent generation and affects the measured PDMR contrast. However, this information was based mainly on fitting the photocurrent as a function of the number of incoming photons and simultaneous measurement of the spin contrast. In this work, we go significantly further as we include the charge carrier recombination on $N_s$ defect as well as $N_s$ photoionisation and the charge carrier drift. This allows us to find a precise dependence of the spin contrast on the $N_s$ concentration along the charge carrier path. For modelling of this behaviour, we varied the number of $N_S$ defects along the spatial axis and probed its effect on the population of NV charge states and on the PDMR spin contrast. We consider the charge states $N_S^0$ and $N_S^+$ in our calculations[17,23]. We assume that in the initial state, there is an equal probability for $N_s$ to be in the $N_S^0$ and $N_S^+$ charge states[17,23], i.e. $p(N_S^0) = 0.5$. The charge state neutrality was compensated by adding a hole with a probability of 0.5 inside the valence band.

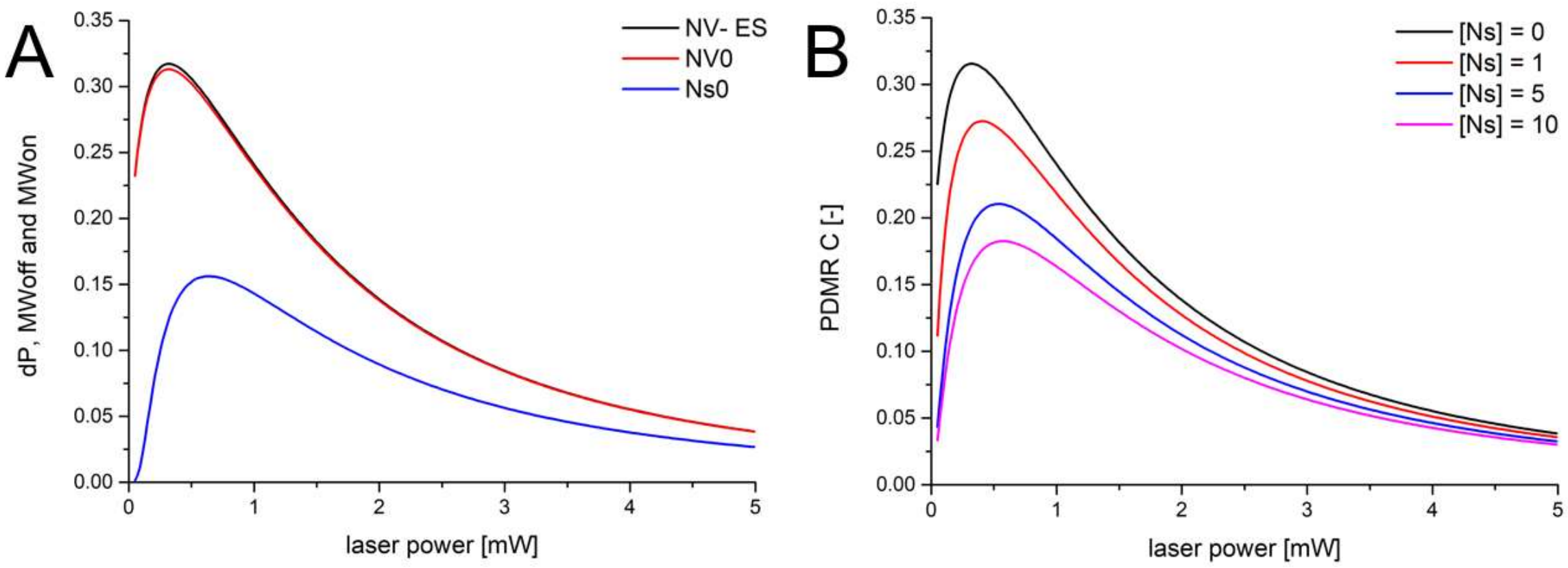

**Figure 6A-B:** Fig 6A shows modelling of the occupation changes (dP) of the $NV^-$ excited state, $NV^0$ and the $N_S$ when the resonant microwave field is turned on and off. In this modelling 1 $N_s^0$ centre and 1 NV centre under a 10 G microwave excitation is considered. Figure 6B shows the laser power dependence of the total PDMR spin contrast for different numbers of $N_S$ present. The total photocurrent (resulting from NV and $N_S$ centres ionisation) is considered here. Traces of different colours represent the output of the model for 1 NV centre and various numbers of $N_S$ centres, ranging from 0 (reference) to 10 (fraction $N_S$/NV of 10/1).

We placed an $N_S$ centre into the central bin of the 1D mesh, i.e. in the same place as the NV defect (see SI Appendix Fig. S3). The total spin contrast is defined by eqn. 12 and is influenced by photoionisation from the $N_S$ level[15]. With our model, we can monitor the changes due to photoionisation and recombination of the charge carriers on individual defect levels. Figure 6A shows the laser power dependence of the occupation change in the total $NV^0$, $NV^-$ excited state and $N_S$ populations when the MW excitation is changed from "on" to "off" (see SI Appendix Eqn. S31).

The calculation is carried out for the situation of 1 Ns centre and 1 NV centre. Notably, the $N_s^0$ occupation is also influenced by microwave application. This is to be expected as $NV^0$ occupation which is involved in the back conversion , Is influenced by the Ns occupation, via charge carrier re-excitation and recombination

Also, the $NV^0$ occupation change is mirroring to the $NV^-$ occupation change, meaning that the changes in the $m_S = 0/\pm1$ can also be potentially monitored by the $NV^0$ population changes, or with less contrast even by $N_S^0$ population change, which brings interesting alternatives for the PDMR detection via defect centres.

Figure 6B depicts the predicted spin contrast on the total photocurrent as a function of the number of Ns centres placed along the spatial axis together with a NV centre. A significant decrease and a shift of the maximum in the electron spin contrast to higher powers is observed when the number of Ns defects introduced to the system increases, as already documented experimentally[15]. As a direct use of the modelling, from the detected shift in the maximum of the PDMR contrast as a function of the laser power, we are able to evaluate how many $N_S$ centres are present in the close vicinity of the NV centre. As the $N_S$ presence has also consequences on the reduction of the $T_2$ and $T_2^*$ times, our model can be useful for evaluating the number of $N_S$ centres in the close vicinity of NV, helping thus devising the origin of the spin decoherence in single NV centres.

To complete the picture, we also model the situation in which an acceptor defect (denoted X) is added to the central bin. This has been partially discussed in our previous work[28]. There, we experimentally observed an inversion of the sign of PDMR resonances (i.e. the formation of positive resonances in the photocurrent) on electron-irradiated CVD samples containing dense NV ensembles[28]. Calculations show that in the presence of acceptor defects $X^+$ with a large electron capture cross-section (approximately 20 times higher than the electron capture cross-section of $N_S^+$ according to modelling results), the formation of positive PDMR resonances can be explained by an increase in the hole photocurrent under the application of a resonant microwave field. Experimental and modelling results, as well as a detailed analysis of this phenomenon, are presented in reference[28].

### 5.3. Comparison between model output and experimental results

We apply the model to the experimentally measured dependence of the optically and photoelectrically detected spin contrasts of a single NV centre. Considering the 20 ppb nitrogen concentration and the focus point volume of roughly 1 um$^3$, we estimate the number of $N_s$ that have an impact on PDMR contrast to be around 10. All of the rates, except for ionisation and recombination, were taken from the literature (see SI Appendix for details). Concerning the rates $k_5$ and $k_6$ there is no consensus at this point on these rates, therefore we apply the model twice – first we consider the rates from Wirtitsch et.al.[13], then the rates from Tetienne et.al.[12], and get the ionisation and recombination rates for both cases (Fig. 7C).

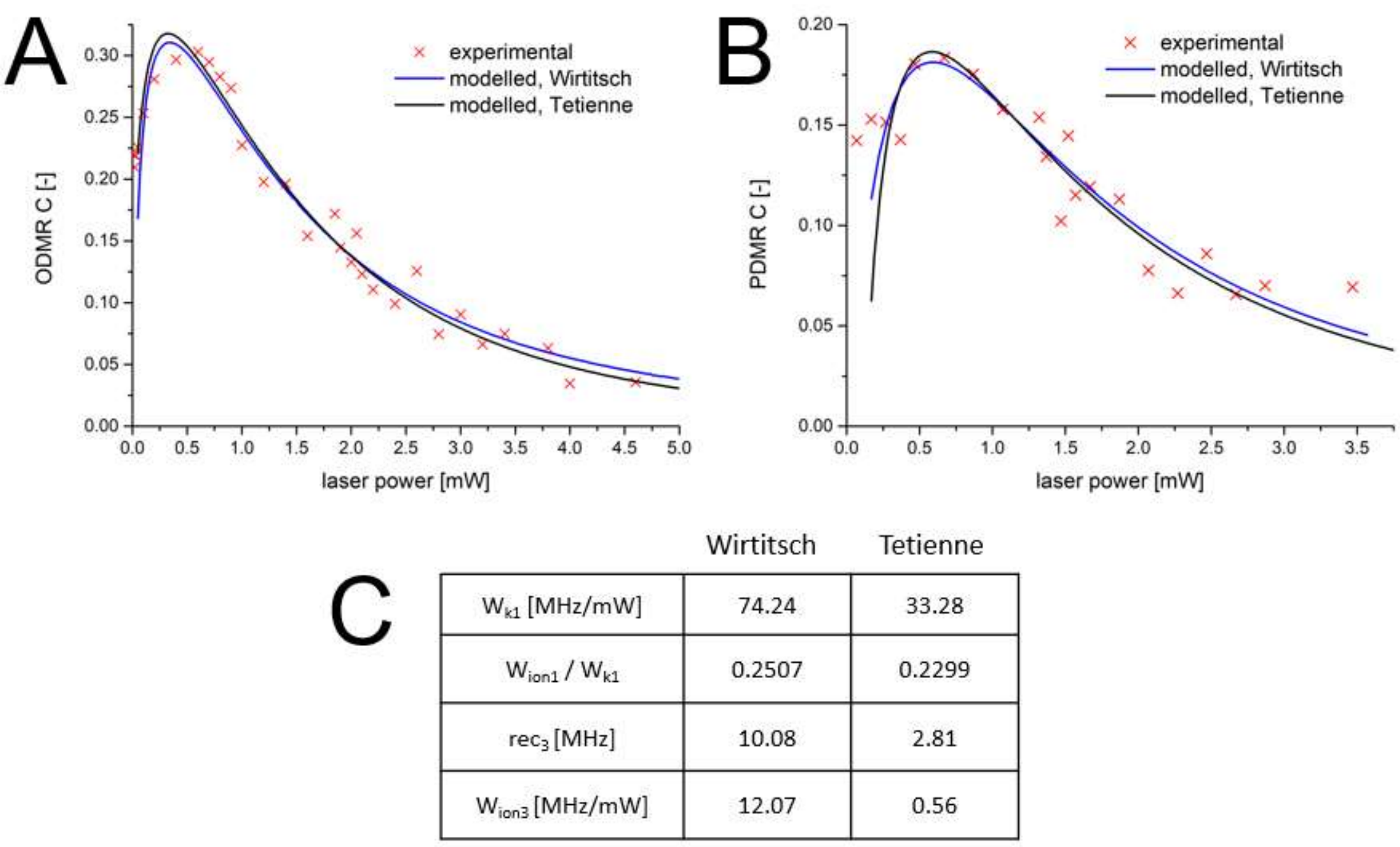

| | Wirtitsch | Tetienne |
|---|---|---|
| $W_{k1}$ [MHz/mW] | 74.24 | 33.28 |
| $W_{ion1}$ / $W_{k1}$ | 0.2507 | 0.2299 |
| $rec_3$ [MHz] | 10.08 | 2.81 |
| $W_{ion3}$ [MHz/mW] | 12.07 | 0.56 |

**Figure 7A-B:** Laser power dependence of the ODMR/PDMR spin contrast for an NV centre in diamond. Fig. 7A shows experimental photoluminescence contrast (red crosses), modelled with Wirtitsch's rates[13] (blue trace) and with Tetienne[12] rates (black trace). Fig. 7B , same as 7A except for photocurrent contrast. 7B leads to an estimated $N_S$ concentration of 10 $N_s$ for 1 measured NV. Fig 7C shows the calculated unknown rates for the two sets of input parameters, either from Tetienne[12] or Wirtitsch[13].

Figure 7 shows a comparison of experimental contrast (red crosses) with the output of the modelled system (blue trace) for both OMDR (Fig. 7A) and PDMR (Fig. 7B). Figure 7A shows that the ODMR contrast peaks at approximately 30 % at about 0.4 mW laser power. Figure 7B shows the experimental and modelled PDMR contrast reduced by Ns defects, where the maximum is approximately 18 % under 1.2 mW laser power (blue trace). Fig 7B also shows a clear significant shift of PDMR contrast maximum that is believed to be caused by the effect of the Ns defect, as shown by modelling.

The ODMR modelling trace for 7A is then recalculated for an identical number of $N_S$ centres. From the modelling we can conclude that, as expected, the ODMR spin contrast is not significantly influenced by the $N_s$ presence, while the PDMR contrast is. From the output of the model (Fig. 7) we can see that recombination rate "$rec_3$" is significantly closer to the previously reported value of 3ns[46] for the case of the input rates from Wirtitsch et.al.

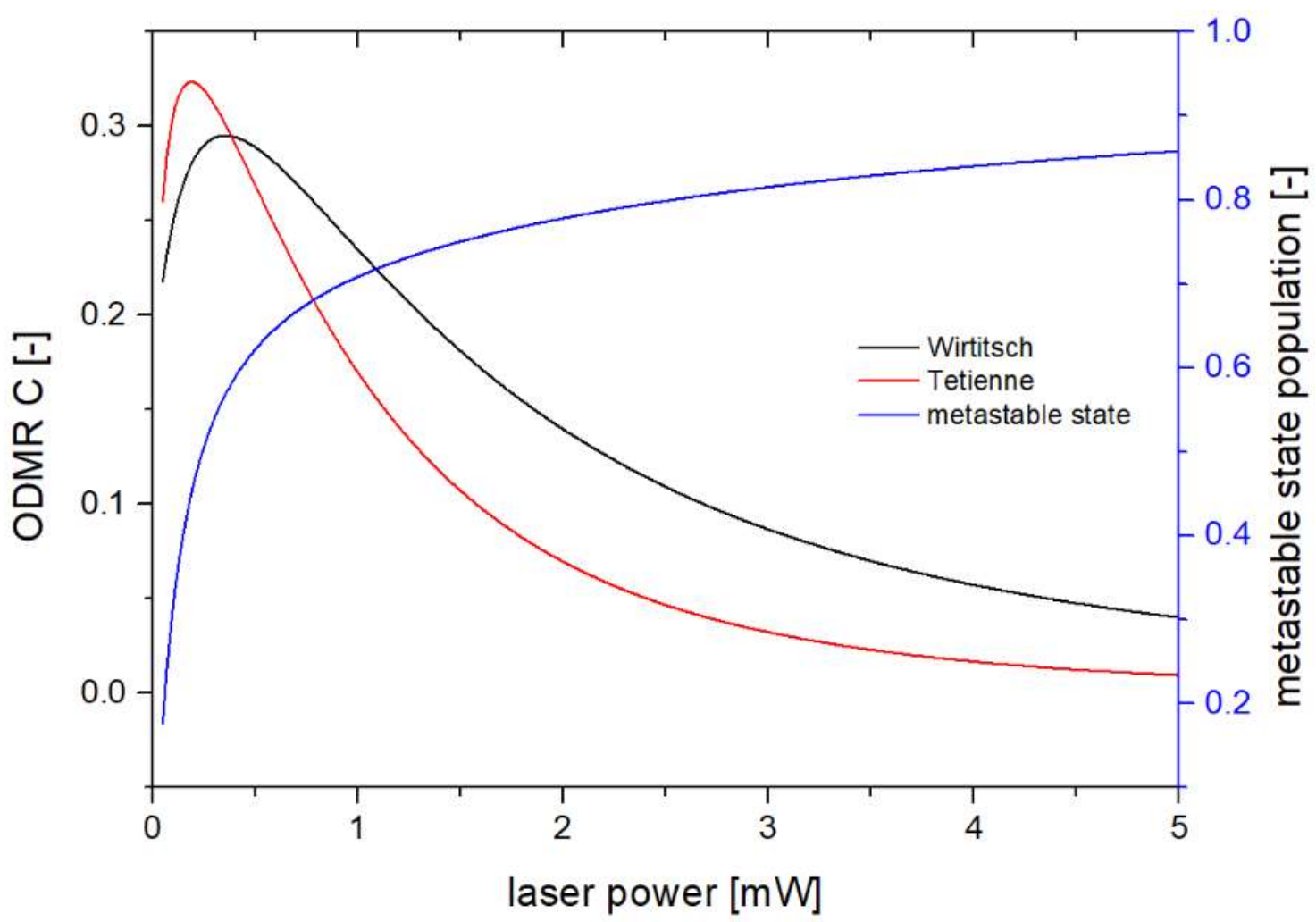

**Figure 8:** Laser power dependence of the ODMR spin contrast in case of 1 NV and 1 $N_s$ defects. Figure shows ODMR contrast with the usage of Wirtitsch's rates (black trace) and Tetienne's rates (red trace) and metastable state population (blue curve).

The modelling of the spin contrast read by ODMR under varying laser powers using the rate constants as defined by Wirtitsch[13] and Tetienne[12] are compared in Fig. 8, for the case of 1 NV centre. A rate for $W_{k1}$ and $W_{ion1}$ was then obtained from fitting of the experimental data with Wirtitsch's rate set and Tetienne's rate set. We see the same trend in modelled traces ( see SI Appendix) ), corresponding to the two different relaxation rates from the metastable singlet state to the $m_s = 0$ state,[12,13]. We show that the results for ODMR spin contrast in Fig. 6B. The two modelled cases (rates from Wirtitsch and Tetienne) differ slightly in the position of the maximal contrast in the PL intensity and contrast is decreasing faster in case of Tetienne's rate set due to the faster dropping of the initialisation efficiency. One of the reason is that the Initialization efficiency, in the case of Tetienne's rate set, depends much stronger on the metastable population, since the transition from the metastable state to the ground state is only weakly selective. In particular, when using the data from Wirtitsch, the ODMR maximum contrast shifts slightly towards the higher laser power.

Notably, by using our theoretical model for the measured ODMR and PDMR spin contrast, we can also calculate the quantum efficiency (QE) for both optical and photoelectric detections. ODMR quantum efficiency has been modelled so far from experimental data[17]. With the fully resolved model, we can calculate theoretically the quantum efficiency and compare it with measured data as a function of the laser power.

Fig. 9 shows the laser power dependence of the fluorescence and photocurrent QE calculated by Eqn. S26-S30. In the low laser power regime, we can see a maximum optical QE of 84 % in the presence of MW and 69% without MW. The optical QE decreases with the laser power. As concerns PDMR, above 2.2 mW, the photocurrent QE increases with the laser power and starts to dominate the optical QE, reaching ~ 44% at 4 mW; however, at this laser power, the photoelectrical spin contrast drops to 10% (see Figure 6B). The results of our model are in a good agreement with the experimental data of QE (that was obtained by fitting eqn. S27). In particular, the PL quantum efficiency calculated using our model is very close to the results presented in reference[15], where the PL QE of the NV centre was measured. In addition to this work, our modelling enables us to obtain the QE for the case of photoionisation. Modelling shows that for higher laser powers, the charge carrier generation becomes more effective than the photon generation by photoluminescence. Despite the decrease in spin contrast, the total photocurrent signal increases with the laser power without saturation, which can be effectively used to reach a higher sensitivity for spin magnetometry. Interestingly, as the photocurrent and PL are complementary, one could extend this measurement concept to measure electron and photon counts correlations, which we plan in the future. Further on, we can see from Fig 9 that the experimental photocurrent QE that is calculated from total collected photocurrent (Eqn. S30) is lower than the theoretical QE trace for 1 NV without $N_S$ defect that is calculated from the generation of charge carriers (Eqn. S28). The decrease of the QE is caused by the fact that the lifetime of the charge carriers is shorter due to the presence of the other defects (Ns). On contrary, if we do the modelling for 10 $N_S$ the modelling matches well the experiment.

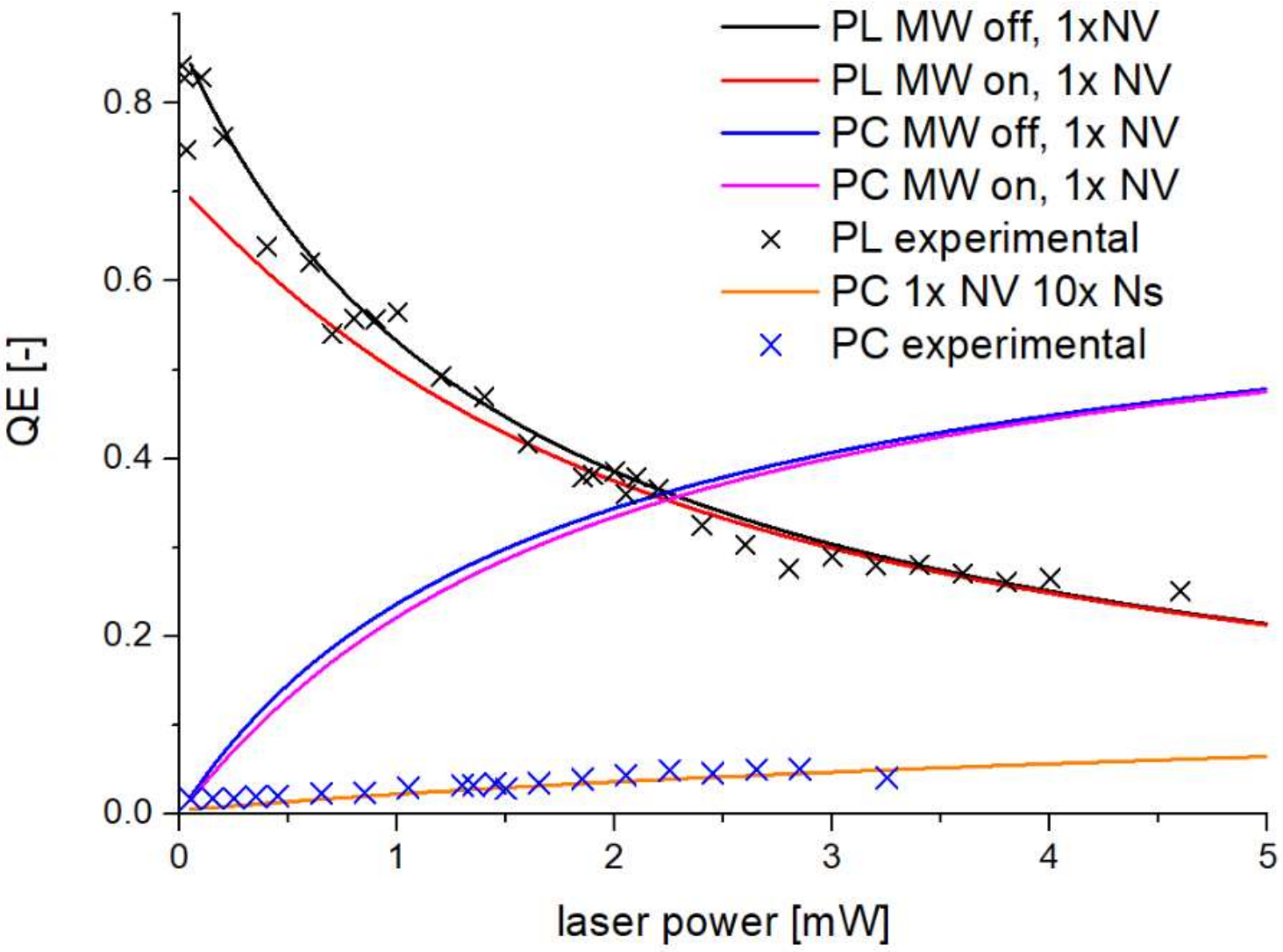

**Figure 9:** Laser power dependence of quantum efficiency. Figure shows the quantum efficiency model of the photoluminescence without (black trace) and with (red trace) microwave excitation for single NV and experimental PL data (black crosses) and the model photocurrent quantum efficiency without (blue trace) and with (pink trace) microwave excitation for 1 NV. The traces are calculated from the PL and photocurrent dependence as a function of the laser power. For the PL data we scale the experimental data onto the theoretical curve to compare the experimental trace shape with the model. The orange trace is the modelled photocurrent quantum efficiency for the case of 1 NV and 10 Ns, and blue crosses correspond to the experimental photocurrent data. The experimental photocurrent data are fitted directly without any scaling.

## 6. Conclusion

In this work, we developed a theoretical model enabling us to describe the occupation of NV energy levels for both $NV^0$ and $NV^-$ charge states. The model is based on a set of rate equations, and contrary to previous models, it incorporates the charge carrier drift obtained from the Boltzmann equation. This enables us to describe the NV electron spin state modulated photocurrent, thus significantly extending previous models, which is especially useful for photocurrent detection of magnetic resonances (PDMR). The model can be used as a guide for optimisation of the experimental photoelectric (but also optical) spin state readout conditions by, for example, setting the optimal laser or MW powers to achieve the highest possible spin contrast. Also, the model includes the presence of different defects, such as substitutional nitrogen $N_S$ (a dominant defect in diamond), and their influence on the spin contrast. The theoretical description allows us to calculate the influence of $N_S$ defects close to the single NV centre that is being monitored from the experimental spin contrast as a function of the laser power. The $N_S$ is an important paramagnetic defect in diamond spin physics as it acts as an electron spin bath, reducing thus the spin dephasing time. Determination of the influence of $N_S$ centres close of the NV centre is therefore of interest. The modelled laser power dependence of the PDMR and ODMR contrasts were compared with experimental data obtained on a single NV defect and showed an excellent agreement. The model also allows the calculation of the quantum efficiency (QE) for photon or charge carrier generation. We could establish that the maximal photon QE (about 84%) is reached at low laser powers, while charge carrier generation is more effective at higher laser powers, exceeding the QE for the photon generation. Additionally, the model allowed us to comment on the value of the rate for the transition from the metastable singlet ground state to the $m_S = 0$ state of the spin-triplet ground state of the NV centre, about which there is an ongoing discussion. Also by modelling of experimental data we could obtain the ratio of the $NV^0$ to NV- charge back conversion rate as well, contributing to the knowledge about the NV centre dynamics under illumination. The proposed model can be considered as an efficient tool for experimenters when optimising the spin state readout in various quantum algorithms for NV centres in diamond, however the set of the equations can be modified to be used for the study of other solid-state spin systems such as SiC or BN.

## 7. Acknowledgements

This work was supported by the Grant Agency of the Czech Republic (Project Number 20-28980S, 24-12984S); EU Horizon 2020 project Amadeus grant agreements ID: 101080136 and No.101046911 QuMicro. The work of M.G. was supported by project no. 101038045 (ChemiQS): This project has received funding from the European Union’s Horizon 2020 research and innovation program. The authors also acknowledge the support from FWO (Funds for Scientific Research Flanders), projects No. G0D1721N and No. G0A0520N.

## 8. Keywords